\documentclass[aps, twocolumn]{revtex4-1}%
\usepackage{amssymb,amsmath,amsthm}
\usepackage{graphicx,color}
\usepackage{epsf}
\usepackage{enumerate}
\usepackage{fancyhdr}
\usepackage{float}

\begin{document}

\author{Alba Ramos and Cecilia Cormick}
\affiliation{Instituto de F\'{i}sica Enrique Gaviola, CONICET and Universidad 
Nacional de C\'{o}rdoba, Ciudad Universitaria, X5016LAE, C\'{o}rdoba, Argentina}

\title{Feasibility of the ion-trap simulation of a class of non-equilibrium 
phase transitions}

\date{\today}

\begin{abstract}
Our work analyzes the potential of ion traps for the experimental simulation 
of non-equilibrium phase transitions observed in certain spin-chain models 
which can be mapped to free-fermion systems. In order to make the dynamics 
more accessible to an experimenter, we first consider relatively small systems, 
with few particles. We analyze phase transitions in the 
non-equilibrium asymptotic regimes of an XY spin chain 
with a transverse magnetic field and coupled to Markovian baths at the end 
sites.
We study a static open system and a case when the spin chain is periodically 
kicked. Notably, in the latter case for some anisotropy parameters the dependence on the system 
size converges rapidly to the many-particle
limit, thus facilitating the experimental observation of the dynamics.
We also define local observables that indicate the 
presence of the quantum phase transitions of interest, and we study the effects 
of the long-range character of the typical interactions obtained in ion traps. 
\end{abstract}

\maketitle

\section{Introduction}

The analysis of non-equilibrium quantum phase transitions 
is a very active area of research (see, for example, \cite{Znidaric_2010, Zoller2011,
Banchi_2014, Heyl_2013, Vosk_2014, Mazza_2012}). The inclusion of driving 
elements, non-thermal baths, and/or baths at different temperatures leads to 
rich phenomena going beyond the traditional quantum phase transitions (QPTs), 
that is, transitions observed in quantum systems at zero temperature and which 
take place as a consequence of quantum rather than thermal fluctuations 
\cite{Sachdev}. The field of non-equilibrium QPTs has seen in the last decades 
a huge 
progress in the possibilities of implementation of the dynamics in the setting 
of controllable quantum systems provided by ultracold atoms \cite{Diehl_2010, 
Baumann_2010}, trapped ions \cite{Jurcevic_2017, Genway_2014}, NMR \cite{Alvarez_2006}, or 
cavity QED \cite{Morrison_2008} to name a few.

A recent series of articles shows the presence of non-equilibrium 
phase-transitions in a particular model of driven XY spin chains, which can be mapped to quadratic 
open fermionic systems 
\cite{Prosen2008, Prosen2010, Prosen2011}. In this work we provide an 
analysis of the feasibility of the implementation of this kind of dynamics in a 
system of trapped ions. This implies the study of several key aspects: in the 
first place, one must translate the elements of the original proposal in terms 
of the dynamics of trapped ions. In this respect, the area of quantum 
simulations in ion traps 
is very advanced and our proposal benefits from the resources already 
available in the literature. We do not elaborate on this aspect since 
the tools to simulate spin systems with ions are standard and described in 
detail elsewhere \cite{Porras_2004, Blatt_2012}. 

More importantly, it is 
necessary to identify the essential 
signatures of the phase transitions that are still visible in chains with 
reduced numbers of particles; even though some research groups have been able to 
perform quantum simulations with large numbers of particles \cite{Zhang_2017, Britton_2012}, most 
trapped-ion 
labs are able to manipulate short chains only. Another difficulty is the fact 
that the original models studied in \cite{Prosen2008, 
Prosen2010, Prosen2011} contain only nearest-neighbour interactions, 
which is essential for the mapping of the 
spin system to a fermionic model admitting an efficient analytical treatment. However, such 
short-ranged interactions are very challenging for an experimenter. We thus focus on power-law 
interactions, which are standard in ion traps but give rise to spin models whose dynamics are harder 
to calculate. 
Finally, it is important to define measurable 
quantities that allow one to observe the phenomena of interest with a 
moderate number of resources and in moderate time; again, the original models focus on 
observables that greatly simplify the theoretical calculation but which are not straightforward to 
measure in an actual implementation because of their non-local character.

This work is organized as follows: In Section \ref{sec:Prosen} we shortly 
review the theoretical models of interest. Section \ref{sec:finite} describes 
the results obtained in the original models when the numbers of chain sites are 
relatively small, and discusses the convergence to the many-particle limit. 
Section \ref{sec:proposal} analyzes modifications of the model in order to make 
it closer to realistic experimental situations: firstly, by changing the 
observable of interest in order to make it more easily accessible to measurements, and 
secondly, by considering the effect of long-range correlations. Finally, in 
Section \ref{sec:conclusions} we summarize our results. Further details are 
provided in three appendices.

\section{The model of interest: an out-of-equilibrium spin chain}
\label{sec:Prosen}

The starting point of our work is provided by two articles by Prosen and 
co-authors \cite{Prosen2008, Prosen2011}. In the first paper, 
a problem of an open and out-of-equilibrium spin chain is presented which 
admits an exact solution of the non-equilibrium steady state using quantization in the space of 
operators (see the Appendix \ref{app:formalism} and \cite{Prosen2008}). For the steady state found, 
a quantum phase transition is detected by
studying long-range correlations. In the second work, a 
similar analysis is carried out in presence of a time-dependent periodic 
driving \cite{Prosen2011}. The authors found complex phase diagrams 
displaying phases of long range and exponentially decaying spin-spin 
correlations depending on the parameter values chosen. For 
the periodically kicked chain, the structure of this phase diagram was related 
with the stationary points of the Floquet quasiparticle dispersion 
relation.

The first model was the time-independent XY chain of $N$ spins $1/2$ with 
transverse field, with Hamiltonian \cite{Lieb_1961, Pfeuty_1970}:
\begin{equation} \label{eq_static}
 H_\text{t-i}=\sum_{m=1}^{N-1}\left(\frac{1+\gamma}{2}\sigma_m^x\sigma_{m+1}^x
 +\frac{1-\gamma}{2}\sigma_m^y\sigma_{m+1}^y\right)
 +h\sum_{m=1}^N \sigma_m^z,
\end{equation}
where $\gamma$ is the anisotropy parameter, $h$ is the magnetic field, and for 
simplicity we set $\hbar =1$.
The system is coupled to a pair of Lindblad baths at its ends, with four
Lindblad operators $L_{1,2}=\sqrt{\Gamma_{1,2}^L}\sigma_1^{\pm}$, $L_{3,4}=
\sqrt{\Gamma_{1,2}^R}\sigma_N^{\pm}$, where $\sigma_j^\pm=(\sigma_j^x\pm 
i\sigma_j^y)/2$. The full evolution of the system's 
density matrix $\rho$ is thus given by
\begin{equation}
 \dot \rho = \mathcal{L}\rho = -i [H,\rho] + \sum_{j=1}^4 2 L_j \rho L_j^\dagger 
- \{L_j^\dagger L_j, \rho\}
\end{equation}
with the curly brackets denoting the anticommutator. As was 
shown in \cite{Lieb_1961, Pfeuty_1970}, the Hamiltonian of the system can be 
diagonalized by mapping the problem onto a system of free fermions. In the 
extended model including the coupling with baths at the ends, the Lindblad 
operators are such that the problem in terms of fermions is still solvable with 
a similar approach \cite{ProsenNJP2008}.

We also consider the periodically kicked XY spin chain governed by the 
Hamiltonian:
\begin{eqnarray}
 H_\text{t-d} (t)&=&H_0+H_1 (t),\\
 H_0&=&\sum_{m=1}^{N-1}\left(\frac{1+\gamma}{2}\sigma_m^x\sigma_{m+1}^x
 +\frac{1-\gamma}{2}\sigma_m^y\sigma_{m+1}^y\right),\\
 H_1&=&a\,\delta_\tau(t)\sum_{m=1}^N \sigma_m^z,
\end{eqnarray}
where $\gamma$ is the anisotropy parameter, $\delta_\tau(t)=\sum_{m \in 
\mathbb{Z}}\delta\left(t-m\tau\right)$ is a periodic Dirac function with period 
$\tau$, and $a=h\tau$ where $h$ is the intensity of the external magnetic field.
This system is also coupled to a pair of Lindblad baths at its ends, defined in 
the same form as in the previous, time-independent model.

In order to solve the problem, the system of $N$ spins is mapped by means of the Wigner-Jordan 
transformation onto a model with $N$ fermionic 
particles with creation and annihilation operators $c_j^\dagger, c_j$ respectively \cite{Lieb_1961, 
Pfeuty_1970}. In this mapping, the spin projection along the $z$ axis in site $j$ is 
directly related with the presence or absence of a fermionic particle in this site, according to 
$\sigma_j^z \equiv 2 c_j^\dagger c_j - 1$. However, the spin projections along directions $x$ and 
$y$ are mapped into non-local observables in terms of the fermionic system. Alternatively, instead 
of working with the $2N$ fermionic creation and annihilation operators, 
one can use $2N$ Majorana operators $\omega_j$ which are given by Hermitean combinations of the 
former. The mapping between spins and Majorana operators is of the form \cite{Prosen2011}:
\begin{eqnarray}
 \omega_{2j-1} &=& \sigma_j^x \prod_{j'<j} \sigma_{j'}^z \,,\\
 \omega_{2j} &=& \sigma_j^y \prod_{j'<j} \sigma_{j'}^z \,.
\end{eqnarray}

The key observable signaling the transitions in the work by Prosen and collaborators was given by 
the 
residual correlation:
 \begin{equation}
  C_{\rm res}=\frac{\left(\sum_{j,k}^{|j-k|\geq N/2}|C_{j,k}|\right)}
 {\left(\sum_{j,k}^{|j-k|\geq N/2}1\right)}.
 \label{eq:Cres}
 \end{equation}
Here, $C_{j,k}$ is the correlation matrix in terms of the Majorana representation of the spin 
operators \cite{Prosen2011}:
\begin{equation}
 C_{j,k} = {\rm tr} (\omega_j \omega_k \rho) - \delta_{j,k} \,.
 \label{eq:Cjk}
\end{equation}
For the case 
of the kicked model, the correlations in Eq.~(\ref{eq:Cres}) are evaluated immediately after the 
kick. With the goal of a possible implementation in mind, it is essential to notice that the 
correlators 
$C_{j,k}$ are local in terms of the fermionic model, but highly nonlocal with respect to the 
original spin operators. This aspect will be analyzed in Section \ref{sec:proposal}.

\section{Numerical results of the original model in small chains}
\label{sec:finite}

We now proceed to numerically study the behaviour of this system in cases with 
few particles. 
We first consider the case of the time-independent non-equilibrium system and 
then we move on to analyze the case of the kicked chain.
As was reported in Refs. \cite{Prosen2008,Prosen2011}, for a sufficiently large 
number $N$ of sites, the main features of the long-range correlation diagrams are no longer 
affected by the details of the baths at the ends. This is not surprising since the strength of 
the coupling to the baths is kept constant while the system is taken to the 
thermodynamic limit; thus, the critical behaviour observed corresponds to the Hamiltonian part of 
the dynamics. However, the presence of some kind of bath is essential 
for the analysis since it is necessary in order to have a well-defined 
asymptotic state. Furthermore, the convergence to this state is fastest when the 
couplings with the baths are taken to be of the same order of magnitude as the Hamiltonian 
parameters.

We focus our study on the feasibility of the implementation with small chains for several reasons: 
first of all, most ion-trap labs in the world can only successfully control small systems. Secondly, 
the transitions are observed in the asymptotic state, and the rate of convergence to it decreases as 
the chains get longer. Finally, the number of observables that one needs to measure in order to 
build the correlation diagram of Prosen and co-authors increases with the number of sites in the 
chain (for more details see Subsection \ref{subsec:local}). Thus, implementations with relatively 
long chains become extremely challenging. In the following we show results concerning the behaviour 
of the different models for an increasing value of $N$; more details are provided in Appendix 
\ref{app:number}.

\subsection{Time-independent model}
\begin{figure}[t]
\begin{center}
\includegraphics[width=0.23\textwidth]{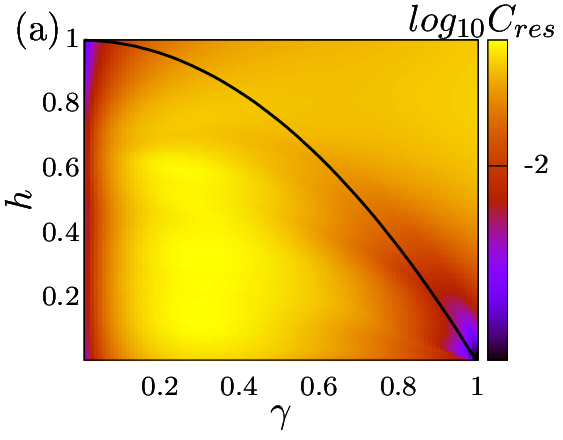}
\includegraphics[width=0.23\textwidth]{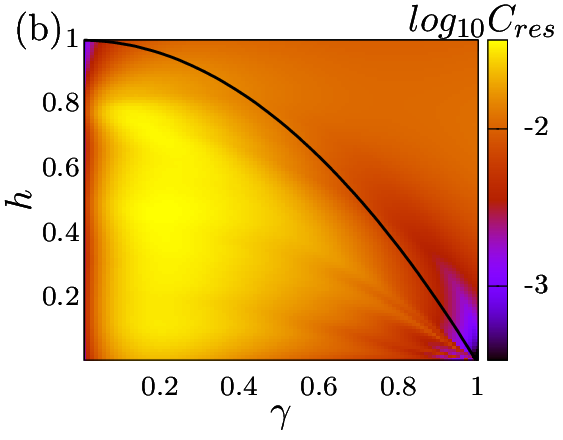}
\includegraphics[width=0.23\textwidth]{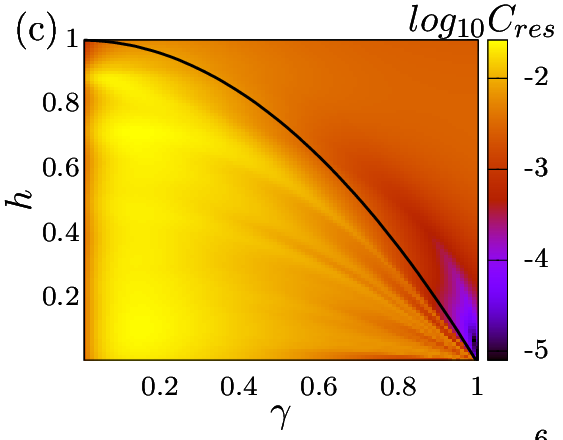}
\includegraphics[width=0.23\textwidth]{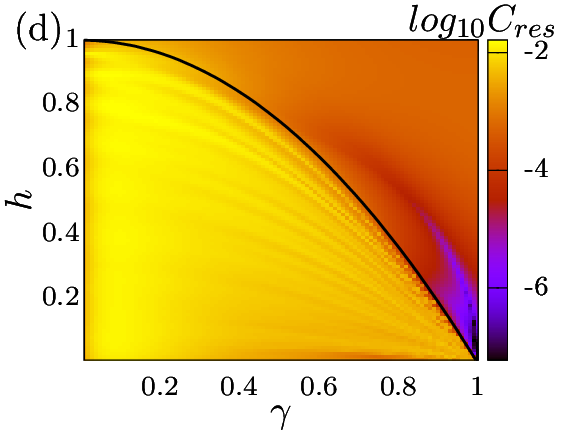}
\end{center}
\caption{\label{fig_8} Diagram in the $\gamma-h$ plane showing residual correlator
$C_{\rm res}$ plotted in a log-scale with $\Gamma_1^L=\Gamma_1^R=0.5$, 
$\Gamma_2^L=0.3$, $\Gamma_2^R=0.1$. The black dashed line shows the 
function $h=1-\gamma^2$ which
determines the critical field $h_c$. Each panel shows the residual correlation diagram for: 
(a)\,$N=5$, (b)\, $N=7$, (c)\, $N=10$ and (d)\,$N=17$.}
\end{figure} 
In order to study the feasibility of the experimental implementation of 
the first model, we analyze the behaviour of the residual correlations 
\eqref{eq:Cres} as a function of the parameters $\gamma,\,h$ and observe the 
convergence to the limit $N\gg1$. As an example, Figure \ref{fig_8} shows the 
residual correlations for a particular example with 
$\Gamma_1^L=\Gamma_1^R=0.5$, $\Gamma_2^L=0.3$, $\Gamma_2^R=0.1$. The different 
subfigures show cases for different particle numbers, $N=$ 5, 7, 10, and 17, 
while the black line indicates the phase boundary in the thermodynamic 
limit. 
The results show that signatures of the QPT appear for $N\gtrsim15$. 
For those values, $C_{\rm res}$ presents an abrupt change which coincides with the 
theoretical prediction $h_c=1-\gamma^2$ \cite{Prosen2008}.

\subsection{Model with periodic kicks}

\begin{figure}[t]
\begin{center}
\includegraphics[width=0.23\textwidth]{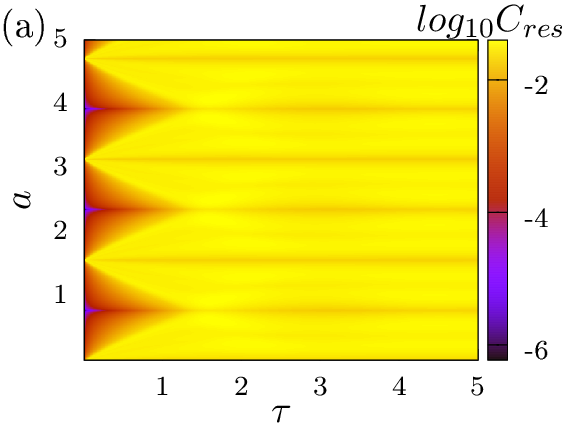}
\includegraphics[width=0.23\textwidth]{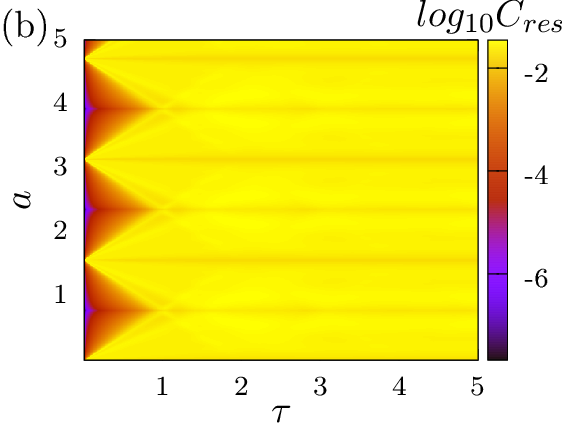}
\includegraphics[width=0.23\textwidth]{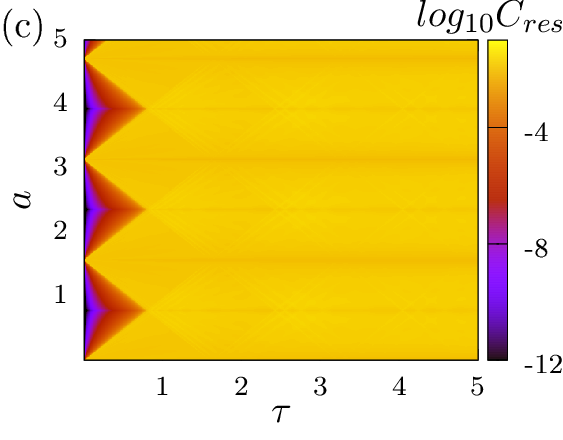}
\includegraphics[width=0.23\textwidth]{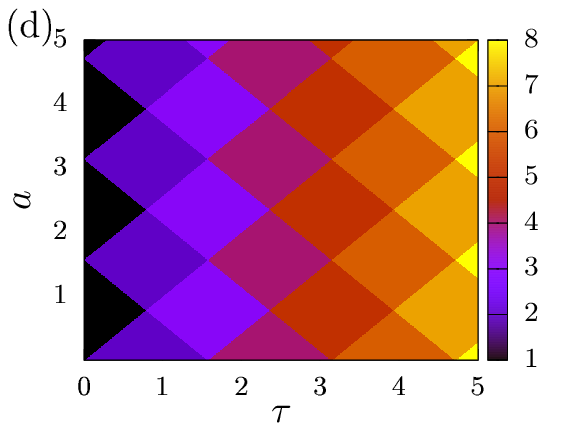}
\end{center}
\caption{\label{fig:gamma01} Residual correlator $C_{\rm res}$ in log scale, in the $a-\tau$ 
plane and with number of sites $N=4$ (a), $N=7$ (b), $N=20$ (c). In panel (d) we show 
half the total number of non-trivial stationary points of the quasi-particle 
dispersion relation for kicked XY chain (see Appendix \ref{app:bands} for details). For all panels 
we used 
$\gamma=0.1$ and the same values of the bath terms that were used in the previous figure.}
\end{figure}

\begin{figure}[t]
\begin{center}
\includegraphics[width=0.23\textwidth]{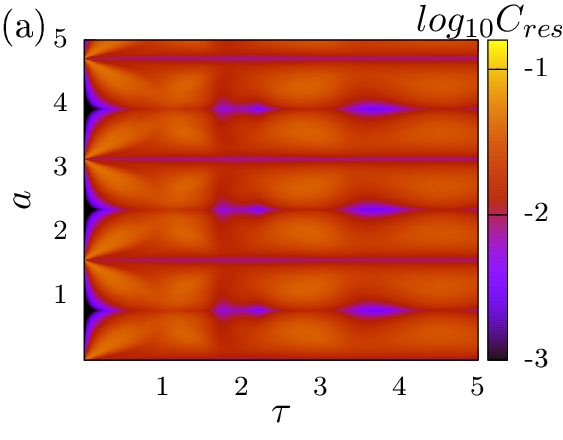}
\includegraphics[width=0.23\textwidth]{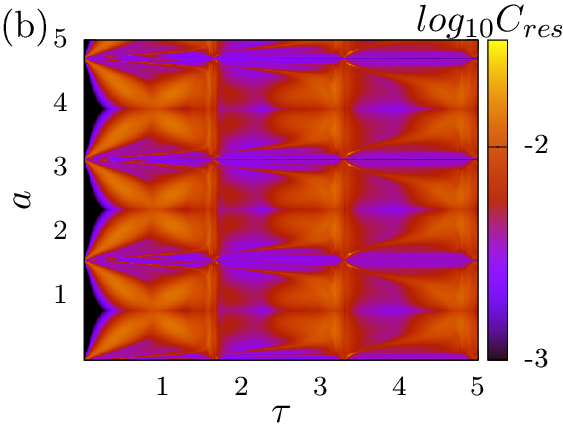}
\includegraphics[width=0.23\textwidth]{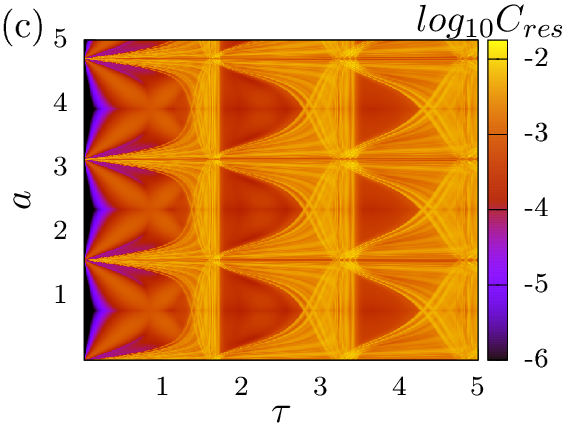}
\includegraphics[width=0.23\textwidth]{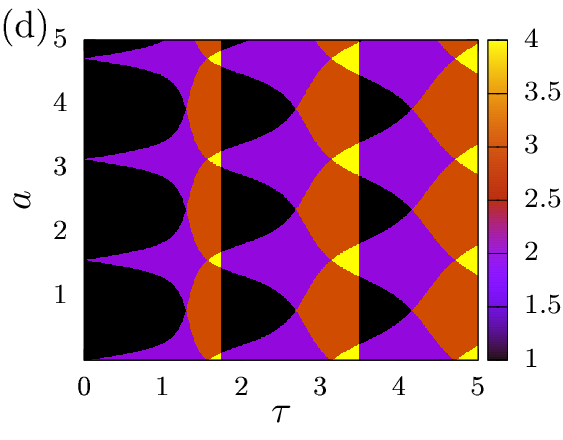}
\end{center}
\caption{\label{fig:gamma09} Residual correlator $C_{\rm res}$ in log scale, in the $a-\tau$ 
plane and with number of sites $N=4$ (a), $N=7$ (b), $N=20$ (c). In panel (d) we show 
half the total number of non-trivial stationary points of the quasi-particle 
dispersion relation for kicked XY chain (see Appendix for details). For all panels we used 
$\gamma=0.9$ and the same values of the bath terms that were used in the previous figures.}
\end{figure}

We now perform a similar analysis of the second model, i.e. the open chain 
with periodic kicks, for different values of the particle number $N$. Figures 
\ref{fig:gamma01} and \ref{fig:gamma09} show the residual correlation \eqref{eq:Cres} 
for $\gamma=0.1$ and 0.9 respectively, with subplots (a-c) corresponding to $N=$ 4, 7, 20. 
We consider the cases $\gamma=0.1$ and $0.9$ as in \cite{Prosen2011}, but we prefer to use 
$a-\tau$ diagrams instead of $h-\tau$ diagrams 
as in \cite{Prosen2011} to display more clearly the periodic dependence of the 
phase diagram on $a$. This periodicity is related with the time-dependent 
driving term  $\sum_m \sigma_m^z$, which has a period  $\frac{\pi}{2}$ and also 
produces a reflection with respect to the axis $a=\frac{\pi}{4}$. Therefore, 
one can restrict the analysis to the interval $a\in [0,\frac{\pi}{4})$.

In panel (d) of Figs. 
\ref{fig:gamma01} and \ref{fig:gamma09}, we show half the total number of 
non-trivial stationary points of the quasi-particle dispersion relation for 
the infinite kicked XY chain (see Appendix \ref{app:bands} and \cite{Prosen2011}). The 
non-equilibrium phase 
transitions correponding to the sudden appearance of long-range correlations happen at points where 
the number of stationary points presents jumps. Thus, this quantity was found to determine the 
shape of the phase diagram \cite{Prosen2011}.

In the case with small 
anisotropy, $\gamma =0.1$, shown in Fig. \ref{fig:gamma01}, one can observe 
that small systems ($N=4$ and 7, in the upper row) already display a structure which resembles that 
of large chains, with a sharp onset of the long-range correlations forming dark triangles on the 
left of the diagram. On the contrary, for large anisotropy parameters, $\gamma=0.9$ as in 
Fig. \ref{fig:gamma09}, one has to reach much larger particle numbers ($N\simeq20$) in order to 
observe a sudden appearance of long-range correlations at 
the points where the number of stationary points exhibits a jump (as displayed in subplot d).
Besides, for $\gamma=0.1$ the transition that is seen most clearly is the one for the 
smallest values of $\tau$. From these observations, in the following we focus on the case 
$\gamma=0.1$ and we restrict to small values of $\tau$.

\section{Towards a realistic experimental implementation}
\label{sec:proposal}

We have previously shown that the time-independent model for 
non-equilibrium phase transitions only displays critical behaviour for 
relatively large chains with $N\gtrsim15$. On the contrary, the model with periodic kicks and 
small anisotropy already shows signatures of the 
phase diagram for very small chains, with $N\gtrsim5$. For this reason, in the 
following we will focus only on this second situation. For definiteness, we will take the number of 
particles fixed at $N=5$.

We consider the possible implementation of the kicked model for non-equilibrium quantum phase 
transitions using a chain of trapped ions. This setup has been proposed, and used, for 
spin simulations in several occasions \cite{Jurcevic_2017, Porras_2004, Blatt_2012, 
Monroe_NJP_2011, Cormick_2013, Jurcevic_2014}. In this context, the internal 
degrees of freedom of the ions encode the spins, while the Coulomb interaction is used to 
generate effective spin-spin interactions. In order to achieve this, ions are driven by laser 
pulses which couple the internal states with the vibrational 
motion of the ions. If the laser pulses are chosen appropriately, an
effective spin-spin interaction is produced \cite{Porras_2004, Blatt_2012}. We will now discuss how 
the 
original model of Prosen and coauthors can be made more experimental-friendly by requiring 
measurements of local 
observables, and after this we analyze the effects of the long range of 
the typical effective interactions achieved in ion-trap simulations of spin 
models. 

\subsection{Local residual correlator} \label{subsec:local}

The residual correlation analyzed so far, Eq.~\eqref{eq:Cres}, is a combination of 
highly non-local
operators in terms of the spin operators \eqref{eq:Cjk}. The experimentally more accessible 
quantities are the two-local correlations \cite{Monroe_NJP_2011, Heyl_2013, Blatt_2012}:
\begin{eqnarray}
 C_{j,k}^{xx}&=&\mbox{tr}\left(\sigma_j^x\sigma_k^x\rho\right)\,,\\
 C_{j,k}^{xy}&=&\mbox{tr}\left(\sigma_j^x\sigma_k^y\rho\right)\,,\\
 C_{j,k}^{yy}&=&\mbox{tr}\left(\sigma_j^y\sigma_k^y\rho\right)\,.
\end{eqnarray}
From them we define the local residual correlation as:
\begin{equation}
 C_{\rm res}^{\rm loc}={\cal P}\left(\sum_{j,k}^{|j-k|\geq 
N/2}\left(|C_{j,k}^{xx}|+|C_{j,k}^{xy}|+
 |C_{j,k}^{yy}|\right)\right)
\end{equation}
where ${\cal P}$ is the normalization factor given by:
\begin{equation}
{\cal P}^{-1} = \sum_{j,k}^{|j-k|\geq N/2} 3 \,.
\end{equation}

\begin{figure}[t]
\begin{center}
\includegraphics[width=0.23\textwidth]{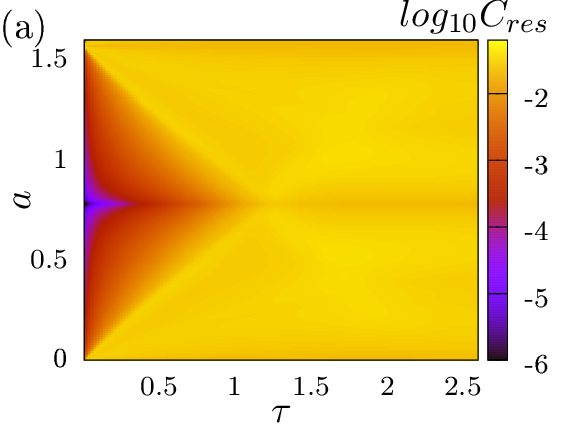} 
\includegraphics[width=0.23\textwidth]{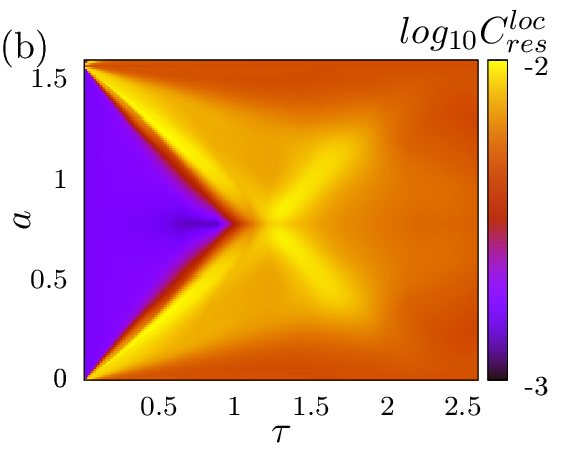}
\end{center}
\caption{\label{fig:local} Residual correlation as a function of $a$ and $\tau$ in log-scale for 
$N=5$ in the representative region $a\in(0,\pi/2]$ and with $\gamma=0.1$.
Panel (a) shows the residual correlation made of observables which are local in the fermionic 
space, panel (b) the one made of observables local in the spin space (see main text for details).
}
\end{figure}

This quantity can be 
numerically computed from the density matrix corresponding to the asymptotic state.
In Fig. \ref{fig:local} we compare local and non-local residual 
correlations for $\gamma=0.1$. We observe that local residual 
correlations show the same structure as the correlations which are local in terms of fermionic 
operators, although with a different intensity (notice the change in the color scale).

We stress that the number of correlations that one needs to measure to obtain $C_{\rm res}^{\rm 
loc}$ is much smaller than the total number of observables necessary for full state tomography. 
However, the amount of measurements required to calculate $C_{\rm res}^{\rm loc}$ still scales up 
as $N^2$, which is another reason to prefer simulations that are feasible 
with small chains. Furthermore, the necessary measurements demand the ability to individually 
address and rotate the desired spins. This is a non-trivial technical requirement, which also 
makes the experiment more challenging for longer chains.

\subsection{Long-range interactions in the ion-trap implementation}

Previous experiments have implemented the  
dynamics of effective Ising models with transverse fields, given by
\begin{equation}
 H_{\rm Ising}= \hbar\sum_{i<k}^N J_{i,k}\sigma_i^x\sigma_k^x
 -\hbar B\sum_i^N\sigma_i^z,
\end{equation}
where $J_{i,k}$ has a dependence on the distance between ions approximately 
given by a power-law decay $J_{i,k}^\eta\propto|j-k|^{-\alpha}$.
The sign of the couplings can be chosen by appropriately tuning the lasers 
\cite{Britton_2012, Monroe_NJP_2011}, and it has been theoretically shown that 
the parameter $\alpha$ can be tuned in the range between 
$\alpha=0$ and $\alpha=3$ \cite{Porras_2004}. Experimentally, tuning of $\alpha$ from 0.01 to 2.72 
has been demonstrated \cite{Britton_2012}. However, interactions in the highest range of 
values of $\alpha$ are hard to achieve experimentally because the large detunings required 
lead to low coupling strengths, and thus most experiments have been performed with intermediate 
values of $\alpha$ \cite{Islam_2013, Jurcevic_2014, Britton_2012, Monroe_NJP_2011}. 

Thus, we now analyze the simulation of our model of non-equilibrium phase transitions in the 
kicked spin chain but where the spin-spin interactions are given by the Hamiltonian:
\begin{equation}
H_{\rm exp}=\sum_{j<k} 
\left(J_{j,k}^x\sigma_j^x\sigma_k^x+J_{j,k}^y\sigma_j^y\sigma_k^y\right),
\end{equation}
where the matrix elements $J_{j,k}^\eta$, with $\eta=x,y$, decay as 
$J_{j,k}^\eta\propto|j-k|^{-\alpha}$.
In our system $J_{j,j+1}^x=\frac{1+\gamma}{2}$ and
$J_{j,j+1}^y=\frac{1-\gamma}{2}$. Considering this Hamiltonian and the term 
associated with the periodic kick we evaluated once more the residual correlations in the 
asymptotic regime. We note that this is numerically more costly, since in presence of long-term 
couplings the Jordan-Wigner transformation does not lead to a quadratic fermionic system any more. 
This means that a compact description of the dynamics is not possible. However, the full 
computation is still feasible due to the small number of particles. 

\begin{figure}[h]
\begin{center}
\includegraphics[width=0.23\textwidth]{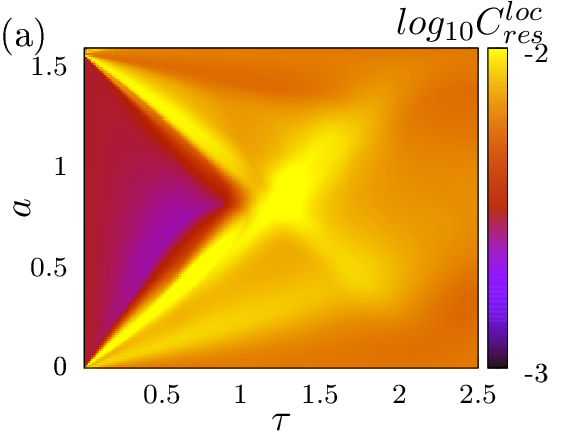}
\includegraphics[width=0.23\textwidth]{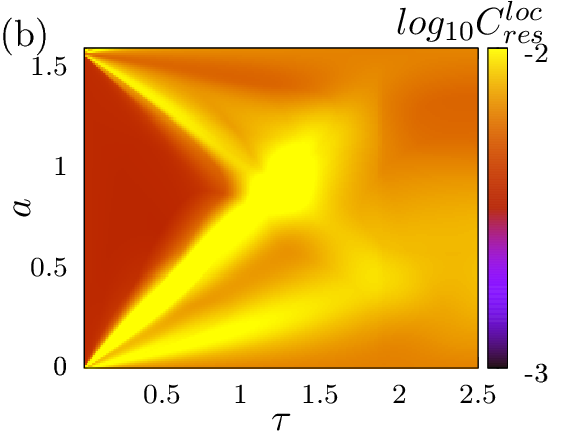}
\end{center}
\caption{\label{fig:long range} 
Local residual correlation maps $a-\tau$ in log-scale for $N=5$
in the representative region $a\in(0,\pi/2]$ and with $\gamma=0.1$.
Here we consider the additional terms associated with the long range of the spin-spin interaction.
We can observe that the reflection symmetry with respect to the axis $a=\pi/4$ is broken; however, 
this effect is small and the diagram still shows a sudden appearance of correlations in 
coincidence with Fig. \ref{fig:local}. For panel (a) a power-law decay with $\alpha=3$ was 
chosen; panel (b) shows the results for the more realistic power $\alpha=2$.}
\end{figure} 

In Fig. \ref{fig:long range} we 
show the results for the local residual correlation $C_{\rm res}^{\rm loc}$ with $\gamma=0.1$ 
including the power-law decay of interactions, with powers $\alpha=3$ (a) and $\alpha=2$ (b).
Both cases are theoretically possible \cite{Porras_2004}; however, case (a), 
which is closest 
to the original model, corresponds to a limiting case of infinite laser detuning. 
Case (b), on the 
contrary, is within the range which has been experimentally demonstrated \cite{Britton_2012}.
We observe that the structure of the diagram is similar to the previous Figures, 
although the new interaction terms affect the reflection symmetry of the correlation 
with respect to the axis $a=\pi/4$. Clearly, the deformation is much more noticeable for the case 
$\alpha=2$ which has stronger interactions for distant pairs of sites; this case also shows a 
lower intensity contrast. Thus, it seems worth pushing 
the interaction power law to the highest possible values ($\alpha=2.72$ being the fastest decay 
achieved so far \cite{Britton_2012}); nevertheless, the qualitative features of the diagram are 
still clearly visible for the case $\alpha=2$.

\section{Conclusions}
\label{sec:conclusions}

We have critically assessed the potential for the ion-trap implementation of the models proposed by 
Prosen \cite{Prosen2008, Prosen2011} displaying non-equilibrium quantum phase transitions in spin 
chains. We analyzed the effects of several key aspects: in the first place, the possibility to 
observe signatures of the phase transitions in chains with small numbers of particles. There, we 
observed that the number of sites required depends on the precise model under consideration: 
for the static model numbers $N\gtrsim15$ are necessary to identify the position of the 
critical line, while for the kicked model the number depends strongly on the anisotropy 
parameter $\gamma$. For $\gamma=0.9$, the shape of the long-range correlation diagram only starts 
ressembling the right critical structure with relatively large numbers, $N\gtrsim20$, whereas a 
small anisotropy $\gamma=0.1$ leads to a diagram displaying the expected onset of 
long-range correlations with a number of sites $N\gtrsim5$. In order to make the necessary number 
of sites as small as possible, we chose to focus on the case $\gamma=0.1$ with small $\tau$.

In a following step, we redefined the original indicator of the presence of 
long-range correlations in order to make it more experimentally accessible. This 
meant replacing 
a quantity which was obtained from local measurements in terms of the fermionic model, and thus 
nonlocal in the experiment, by one which was defined from observables which are local in 
terms of the spin operators. We then concluded that the structure of the diagram was still visible 
when performing this kind of measurement, although the intensity pattern varied with respect to 
the original results. 

Finally, we considered the effect of the presence of long-range interactions 
in the experimentally feasible ion-trap simulations. We assumed that the effective spin-spin 
interactions had a power-law decay, and computed the 
resulting correlation diagrams for powers $\alpha=3$ and $\alpha=2$. While the 
long-range interactions partially destroyed one of the symmetries of the 
original model, one could still observe transitions between regimes with and without long-range 
correlations. We thus conclude that although the exact features 
of the original article \cite{Prosen2011} are not preserved in a realistic implementation, the 
markers of the transitions are still visible in experimentally achievable conditions.

\section*{Acknowledgments}

C.C. acknowledges funding from grant BID-PICT 2015-2236, and helpful discussions with Leonardo 
Ermann about numerical problems.

\appendix
\section{Treatment of Markovian open quasi-free fermionic systems} \label{app:formalism}

We review here the formulation used by Prosen for 
quadratic systems with a bath with linear terms \cite{Prosen2010, Prosen2011}. 
The aim of this formulation is to calculate the correlations avoiding the 
calculation of the  density matrix, and it is a generalization of the standard 
description of Gaussian states by means of the covariance matrix \cite{Gaussian}. This 
approach has considerable advantages, leading to a substantial decrease 
of computational costs in cases when a full analytical solution is not possible.
Within this formalism, correlations are obtained by solving a
Lyapunov equation, for which there are fast computational solving softwares.

The formulation is performed in terms of the Majorana operators.
The authors consider a system of fermions described by
$2N$ anticonmmuting Hermitean operators $\omega_j$, $j=1,...,2N$, 
$\{\omega_j,\omega_k\}=2\delta_{j,k}$. 
One can define Majorana operators in terms of Pauli matrices on $N$ sites as
\begin{eqnarray}
 \omega_{2m-1}&=&\sigma_m^x\prod_{m'<m}\sigma_{m'}^z,\nonumber\\ 
\omega_{2m}&=&\sigma_m^y\prod_{m'<m}\sigma_{m'}^z,
\end{eqnarray}
The authors consider systems for which both the Hamiltonian ${\cal H}$ and the 
Lindblad operators $L$ can be simultaneously expressed in
terms of a quadratic form and linear forms respectively. In this representation
\begin{equation}
 {\cal H}=\sum_{j,k}\omega_j 
H_{j,k}\omega_k\equiv\underline{\omega}\cdot\mathbb{H}
 \underline{\omega},
\end{equation}

\begin{equation}
L_\mu=\sum_j\ell_{\mu,j}\omega_j\equiv\underline{\ell}_\mu\cdot\underline{\omega
},
\end{equation}
where $\mathbb{H}$ can be chosen to be antisymmetric and imaginary, and 
$\underline{\ell}_\mu\in\mathbb{C}^{2N}$.

The object of interest is the correlation matrix ${\mathbb C}$ with elements $C_{j,k}$ defined as:
\begin{equation}
 C_{j,k}=\mbox{tr}\left(\omega_j\omega_k\rho\right)-\delta_{j,k}
\end{equation}
If the equations for the evolution of the system do not depend on time, one can 
obtain the elements of $\mathbb{C}$ in the stationary state through the continuous Lyapunov
equation \cite{Prosen2010}:
\begin{equation}
 \mathbb{XC+CX}^T=i\mathbb{Y}\,.
\end{equation}
Here, $\mathbb{X}=4\left(i\mathbb{H}+\mathbb{M}_r\right)$ and $\mathbb{Y}=
4\left(\mathbb{M}_i-
\mathbb{M}_i^T\right)$, with the real matrices 
$\mathbb{M}_r,~\mathbb{M}_i$ defined from $\mathbb{M}=\mathbb{M}_r+i\mathbb{M}_i$,
$\mathbb{M}=\sum_\mu\underline{\ell}_\mu\otimes\overline{\underline{\ell}}_\mu$.

On the other hand, when the evolution equations are time-dependent, the 
covariance matrix $\mathbb{C}$ satisfies
\begin{equation}\label{eq_cov_t}
 \dot{\mathbb{C}}(t)=-\mathbb{X}(t)\mathbb{C}(t)-\mathbb{C}(t)\mathbb{X}^T(t)+i\mathbb{Y}(t)
\end{equation}
with $\mathbb{X}(t)$ and $\mathbb{Y}(t)$ defined as before but from the time-dependent $\mathbb{H}$ 
and $\mathbb{M}$.
The solution to eq. (\ref{eq_cov_t}) can be written as \cite{Prosen2011}
\begin{equation} 
\mathbb{C}(t)=\mathbb{Q}(t)\mathbb{C}(0)\mathbb{Q}^T(t)-i\mathbb{P}(t)\mathbb{Q}
^T(t),
\end{equation}
where $\mathbb{P}(t)$ and $\mathbb{Q}(t)$ have to fulfill
\begin{eqnarray}
 \dot{\mathbb{Q}}(t)&=&-\mathbb{X}(t)\mathbb{Q}(t),\\
 \dot{\mathbb{P}}(t)&=&-\mathbb{X}(t)\mathbb{P}(t)-\mathbb{Y}(t)[\mathbb{Q}^T(t)]^{-1}
\end{eqnarray}
with the initial conditions $\mathbb{Q}(0)=\mathbb{I},~\mathbb{P}(0)=0$.

In the case studied in \cite{Prosen2011}, the temporal dependence of the problem is 
through a periodic kick in the Hamiltonian, so that
${\cal L}(t+\tau) ={\cal L}(t)$, where $\tau$ is the period. The correlation 
matrix for the stationary state is then obtained solving the discrete Lyapunov 
equation
\begin{equation}
 \mathbb{Q}(\tau)\mathbb{C}_F\mathbb{Q}^T(\tau)-\mathbb{C}_F=i\mathbb{P}(\tau)
 \mathbb{Q}^T(\tau),
\end{equation}
where $\mathbb{C}(0)=\mathbb{C}(\tau)=\mathbb{C}_F$.

\section{Quasi-energy bands of the kicked system} \label{app:bands}

In Ref. \cite{Prosen2011}, analyzing the kicked system, it was shown that,
when dealing with many particles, the main properties of the residual correlations
depend essentially on the Hamiltonian of the bulk.
For this reason, the dispersion relation of a kicked XY infinite chain was studied.
The system has two quasi-energy bands of the form:
\begin{eqnarray}
 \theta_{1,2}(\kappa)&=&\pm \mbox{arcos}\Big\{\cos\left(2\tau h\right)
 \cos\left[2\tau\varepsilon(\kappa)\right]\nonumber\\
 &&+\sin\left(2\tau h\right)
 \sin\left[2\tau \varepsilon(\kappa)\right]
 \frac{\cos(\kappa)}{\varepsilon(\kappa)}\Big\}
\end{eqnarray}
where $\varepsilon(\kappa)=\sqrt{\cos^2(\kappa)+\gamma^2\sin^2(\kappa)}$ is the
quasi-particle energy for the unkicked XY chain, 
$\kappa\in\left[\left.-\pi,\pi\right.\right)$.
As discussed in Refs. \cite{Prosen2011,Prosen2010} the jumps in the residual 
correlations $C_{\rm res}$ coincide
with jumps in the number of non-trivial stationary points 
of the quasienergies dispersion relation of the Floquet eigenstates of the 
system. As the total number of stationary points is always even, the quantity shown in the plots 
in the main text is half this number.

\section{Dependence of the residual correlations on the number of sites}
\label{app:number}

\begin{figure}[h]
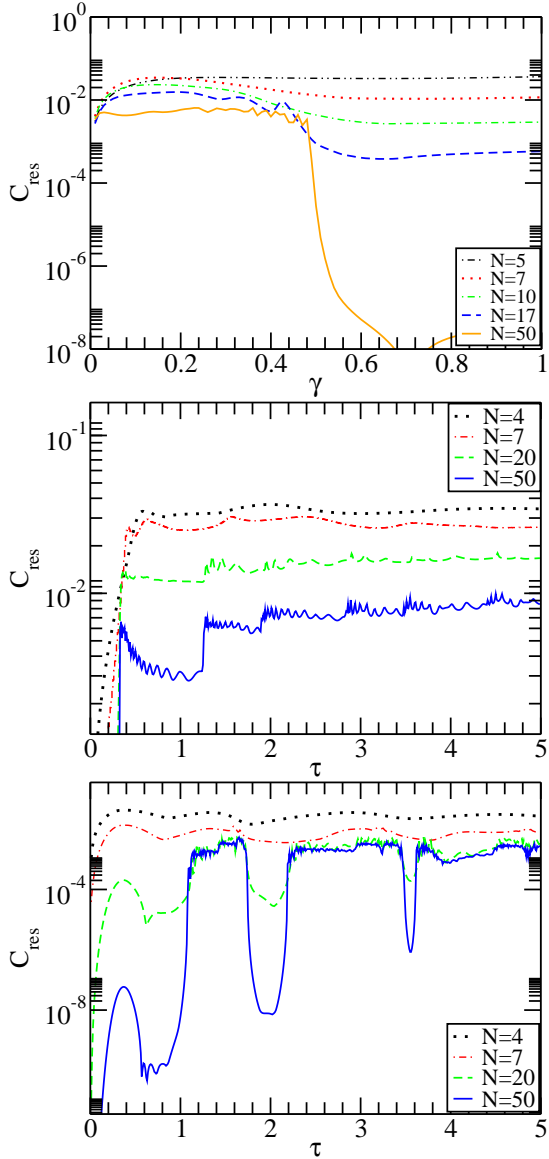

\begin{center}
\includegraphics[width=0.4\textwidth]{h0p75_tau_estatico.eps}
\includegraphics[width=0.4\textwidth]{a1p25_tau_gamma0p1.eps}
\includegraphics[width=0.4\textwidth]{a1p25_tau_gamma0p9.eps}
\end{center}
\caption{\label{fig_cuts} Residual correlator
$C_{\rm res}$ plotted in a log-scale with $\Gamma_1^L=\Gamma_1^R=0.5$, 
$\Gamma_2^L=0.3$, $\Gamma_2^R=0.1$, and for different values of $N$.  Plot (a) corresponds to the 
static case as a function of $\gamma$, with $h=0.75$; plots (b) and (c) correspond to the kicked 
model and show $C_{\rm res}$ as a function of $\tau$, for $\gamma=0.1$ and $0.9$ respectively, 
and with $a=1.25$.}
\end{figure}

In this Appendix we provide further plots showing the behaviour of the residual correlation 
$C_{\rm res}$, defined in Eq.~(\ref{eq:Cres}), as the number of sites is increased. In Fig. 
\ref{fig_cuts} we 
plot the residual correlation for different values of $N$. Plot (a) corresponds to the static case 
as a function of $\gamma$, with $h=0.75$, whereas plots (b) and (c) correspond to the kicked model 
and show $C_{\rm res}$ as a function of $\tau$, 
with $a=1.25$, for $\gamma=0.1$ and $0.9$ respectively. One can 
see that in plots (a) and (c) a number of sites $N$ around 20 is necessary to observe the signatures 
of the transition. On the contrary, plot (b) indicates that for small anisotropy, $\gamma=0.1$, the 
transition at smallest $\tau$ is already visible with less than 10 sites.

\end{document}